\documentclass[12pt,a4paper]{article}
\usepackage[]{epsfig}

\begin{document}

{\LARGE \begin{center}
Limits on WIMP dark matter using scintillating CaWO$_4$ cryogenic 
detectors with active background suppression
\end{center}}

\begin{center}
G.\,Angloher $^a$, 
C.\,Bucci $^d$, 
P.\,Christ $^a$,
C.\,Cozzini $^b$, 
F.\,von Feilitzsch $^c$, 
D.\,Hauff $^a$, 
S.\,Henry $^b$, 
Th.\,Jagemann $^c$, 
J.\,Jochum $^e$, 
H.\,Kraus $^b$, 
B.\,Majorovits $^b$, 
J.\,Ninkovic $^a$, 
F.\,Petricca $^a$, 
W.\,Potzel $^c$, 
F.\,Pr\"obst $^{a,~+}$, 
Y.\,Ramachers $^{b,~*}$, 
M.\,Razeti $^c$, W.\,Rau $^c$, 
W.\,Seidel $^a$, 
M.\,Stark $^c$, 
L.\,Stodolsky $^a$, 
A. J. B. Tolhurst $^b$, 
W.\,Westphal $^c$, 
H.\,Wulandari $^c$ 
\end{center}
 
\centerline{\it\small $^a$ MPI f\"ur Physik, F\"ohringer Ring 6, 
80805 Munich, Germany}
\centerline{\it\small $^b$  University of
Oxford,  Department of Physics,
Oxford OX1 3RH, U.K.}
\centerline{\it\small $^c$ Technische Universit\"at M\"unchen,
Physik Department, D-85747 Garching, Germany}
\centerline{\it\small $^d$ Laboratori Nazionali del Gran Sasso,
I-67010 Assergi, Italy }
\centerline{\it\small $^e$ Eberhard-Karls-Universit\"at T\"ubingen,
D-72076 T\"ubingen, Germany}
\centerline{\it\small $^+$ corresponding author. Tel.:
+49-89-32354-270.
E-mail address: proebst@mppmu.mpg.de}
\centerline{\it\small $^*$ present address: University of Warwick,
Coventry CV4 7AL,
U.K. }

PACS numbers: 95.35.+d, 29.40.-n, 29.40.mc
WIMPs; dark matter; low temperature detectors; CaWO$_4$; background
discrimination

\begin{abstract}
We present first significant limits on WIMP dark matter by the
phonon-light technique, where
 combined phonon and light signals from a scintillating
cryogenic detector are used. Data from early 2004 with two
300\,g
CRESST-II prototype detector modules   are presented, with  a net
exposure of 20.5\,kg days. The modules consist of a 
CaWO$_4$
scintillating ``target''
crystal and a smaller cryogenic light detector. The
combination  of phonon and light signals 
leads to a strong suppression of non-nuclear recoil  
backgrounds. Using  this information to define an acceptance
region for nuclear recoils we have 16 events from the two modules,
corresponding to a rate  for nuclear recoils between 12 and 40~keV
of  ($0.87\pm 0.22$) events/(kg day). This is  compatible with the
rate
expected from neutron background, and most of these events lie in
the region of the phonon-light plane anticipated for
neutron-induced recoils. A particularly strong limit for WIMPs with
coherent scattering
results from selecting a region of
the phonon-light plane corresponding to tungsten
recoils, where the best module shows zero events.

\end{abstract}
\newpage

\section{Introduction}

Despite persuasive indirect evidence for the existence of dark
matter in the universe and in galaxies, the direct detection of
dark matter remains one of the outstanding experimental challenges
of present-day physics and cosmology.
 
 A plausible candidate for the dark matter is the 
Weakly Interacting Massive Particle (WIMP) and it is possible that
it can be detected by laboratory experiments, particularly using
cryogenic methods, which are well adapted to the small energy
deposit
anticipated~\cite{Goodman}. 
Supersymmetry provides
a well motivated WIMP candidate in the form of the lightest
supersymmetric
particle. WIMPs  \cite{Jungman} are expected to be gravitationally
bound in a
roughly isothermal halo around the visible part of our galaxy with
a density of about 0.3\,GeV/cm$^3$ at the position of the
Earth~\cite{rho}.

Interaction with ordinary matter is expected via elastic
scattering on nuclei. 
 The elastic
nuclear scattering  can occur via coherent (``spin independent'')
or
spin-dependent interactions. For the  coherent case,  a factor 
$A^2$ is expected in
the cross section, favoring heavy nuclei~\cite{coh}. We
present our WIMP limits in terms of this possibility. 

In  the CRESST experiment we attempt to
detect the WIMP-nucleus   scattering  using cryogenic methods.
 Results from the first phase of  CRESST using sapphire ($Al_2O_3$)
detectors have
been previously reported, see ref.~\cite{CRESST-I}.
 For the second  phase~\cite{proposal} presently in
preparation,
CRESST-II,  we have  developed cryogenic detectors based on 
scintillating
CaWO$_4$ crystals. When supplemented with a light detector these
provide very efficient discrimination of
nuclear recoils from radioactive $\gamma$ and $\beta $ backgrounds,
down to recoil energies of about 10~keV. 
The heaviest 
nucleus in our crystal is tungsten, for which 
the recoil energy is expected to reach up to about 40~keV, with 
 rates below 1 event/(kg day).
The mass of each crystal
is about 300\,g.  
 Data were accumulated  with two of these prototype detector
modules
for  53 days  starting January 31 2004 at the
Laboratori Nazionali del
Gran Sasso (LNGS).

 Passive background suppression is achieved with a low
background
installation and the  deep underground  location. The overburden of
3500 meter water equivalent at the LNGS leads to a reduction of the
surface
muon flux by $10^6$ to about 
1/(m$^2$ h), while  the  detectors themselves are shielded against
ambient
radioactivity by thicknesses of 14
cm  of low background copper and 20 cm of low background lead. A
neutron shield and a muon veto, to be installed for CRESST-II, 
were not present for the data presented here.

 The  setup for the run reported here is the same as in
 CRESST-I, described in more detail in
ref. \cite{CRESST-I}.
A  four channel SQUID 
system allowed the simultaneous operation of two phonon/light
modules.   The
measurements  were halted on March 23 2004 to proceed with the 
upgrade, which in addition to the 
 neutron shield and the  muon veto, will involve a 66 channel SQUID
readout system to enable operation of 33 detector modules
\cite{proposal}.

\section{Detectors}

A single detector module consists of a scintillating CaWO$_4$
``target''
crystal, operated as a cryogenic calorimeter (the ``phonon
channel''),
and a nearby but separate cryogenic detector optimized for the
detection of
scintillation photons (the ``light channel'').  The phonon channel
is  designed to  measure the energy transfer to a nucleus
of the
CaWO$_4$
crystal in a WIMP-nucleus elastic scattering. Since a nucleus
 and an  electron or gamma of the  same energy differ
substantially in the yield of scintillation
light, an effective background discrimination against gammas and
electrons is obtained by a
simultaneous measurement of the  phonon and light signals. Among
different
scintillating crystals, CaWO$_4$ was selected because of its
high light yield  at low temperatures and the
absence of a noticeable degradation of the light yield for events
near
the crystal surface \cite{Meunier}. Such a degradation, often found
in coincident 
phonon-charge measurements and some scintillators, can cause
difficulties
as it may lead to a
misidentification of electron/photon surface events as nuclear
recoils.  In addition, the
large atomic mass (A=183.86) of tungsten makes CaWO$_4$ a very
favorable target for WIMPs with coherent interactions. 

\begin{figure}[tbh]
\begin{center}
\epsfig{file=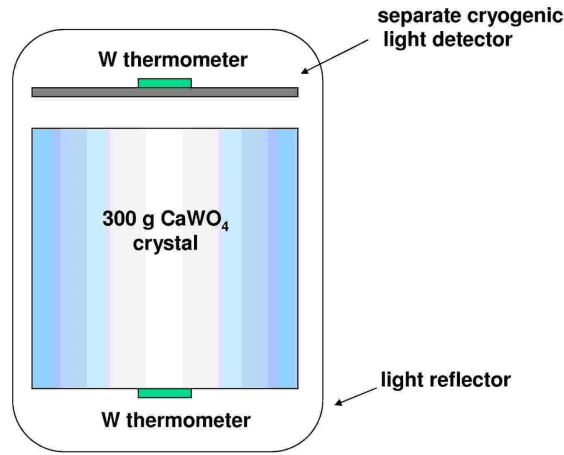,height=12cm,angle=-90}
\end{center}
\caption[]{Schematic representation of the detector for
coincident phonon and light
measurement. It consists of two cryogenic detectors enclosed
in a 
highly reflective housing, read out by tungsten superconducting
phase transition
thermometers.
}
\label{fig:detector_scheme}
\end{figure}

The prototype detector modules  used here \cite{Meunier, Philippe,
LTD10} consist of a 300\,g
cylindrical CaWO$_4$ crystal with 40\,mm diameter and height, and
an
associated cryogenic light detector. The light detector is mounted
close to a flat surface of the
CaWO$_4$ crystal, and both detectors are enclosed in a 
housing  of  highly reflective polymeric multilayer foil
 \cite{3M}.  An
important improvement of the energy resolution of the 
light channel was achieved by
roughening the flat surface of the CaWO$_4$  facing the light
detector to a roughness of about 10\,$\mu$m. This reduces internal
total reflection and  facilitates the
escape of the scintillation photons to the light detector.  Both
detectors are
read
out by optimized tungsten superconducting phase transition
thermometers. The
arrangement is shown schematically in
Fig.~\ref{fig:detector_scheme}. 
The
polymeric foil was measured to have a reflectivity of about 99\,\%
at
the wave
length of the scintillation maximum (420 nm) of CaWO$_4$.  In
addition we have discovered that the carrier foil onto which the 
reflecting foil is grown is itself scintillating at low
temperatures. 
We mounted the reflector with the carrier foil inside, facing
the detectors. This is very useful in
discriminating backgrounds from $\alpha$ decays on the crystal
or
reflector surfaces  where the $\alpha$ particle goes away from the
crystal and only the recoil nucleus enters the
CaWO$_4$ crystal, with little light emission.   The additional
light from the $\alpha$ particle as it is absorbed on the facing
surface of the foil allows the rejection of such events.

An early problem of no-light events induced by stress on the
crystal 
was eliminated by replacing Teflon supports with copper beryllium
springs 
to hold the crystal. These springs were silver coated and other
elements for 
holding the reflecting foil were made of sintered Teflon.
Both
materials are highly reflective in the wavelength range of
interest,
but do not scintillate. 
In future designs these elements will be coated with
an organic scintillator.

The two 
 CaWO$_4$ crystals  used in this work were called ``Julia'' and
``Daisy'', and the associated light detectors were denoted as
``BE 14'' and ``BE 13'' respectively.
The detectors are operated at a temperature of about 10\,mK where
the tungsten thermometer
is in the middle of its transition between the superconducting and
the normal conducting state, so that a small temperature rise of
the
thermometer
leads to a relatively large rise of its resistance. 

The thermal
response of the thermometer film
  can be described by a model \cite{model} which leads to a
pulse described by a rise time and two decay times. Roughly
speaking, the
  initial high frequency phonons created by a particle
 or  photon  interaction  are
 reflected many times at the free
crystal
surface to form a ``hot'' (relative to the initial temperature)
``gas'' of phonons. These athermal phonons are readily absorbed by
the electrons of 
the thermometer. However, after thermalization of their energy in
the metal the
resulting thermal phonons  are very weakly coupled
back to the crystal. This is a kind of ``greenhouse effect'' due to
the strongly decreasing thermal electron-phonon coupling at low
temperature, $g_{e-p} \propto
T^5$.  This effective thermal decoupling of the crystal is
 important for the ``thermal tuning'' of the detectors in that it
permits the relaxation time of the thermometer to be
determined essentially by the heat capacity of the thermometer and
the thermal conductance to the heat bath.
   The thermal time
constant of the thermometer is thus adjusted via a normal
conducting
metallic 
link to the copper crystal holder, which  plays the role of  the
``heat
bath''.

\subsection{Phonon channel} 

The thermometer of the CaWO$_4$ crystals is a 6$\times$8~mm$^2$,
200\,nm thick superconducting tungsten film evaporated on the
surface. In addition there is a buffer layer of SiO$_2$
between the tungsten film and the CaWO$_4$  to prevent
inter-diffusion between the  tungsten film and the CaWO$_4$
crystal.  The quite low superconducting transition
temperatures of 7\,mK and 9\,mK  for the
 present CaWO$_4$  detectors were achieved using the  buffer layer
and a deposition
temperature of 480\,$^0$C.  For the 
tungsten
etching, a dilute
mixture of NaH$_2$PO$_4$, NaOH, and Na$_3$Fe(CN)$_6$ was used
instead of conventional potassium based
etchants to avoid
radioactive contamination by $^{40}K$.
\begin{figure}[tb]
\vspace*{-3cm}
\begin{center}
\epsfig{file=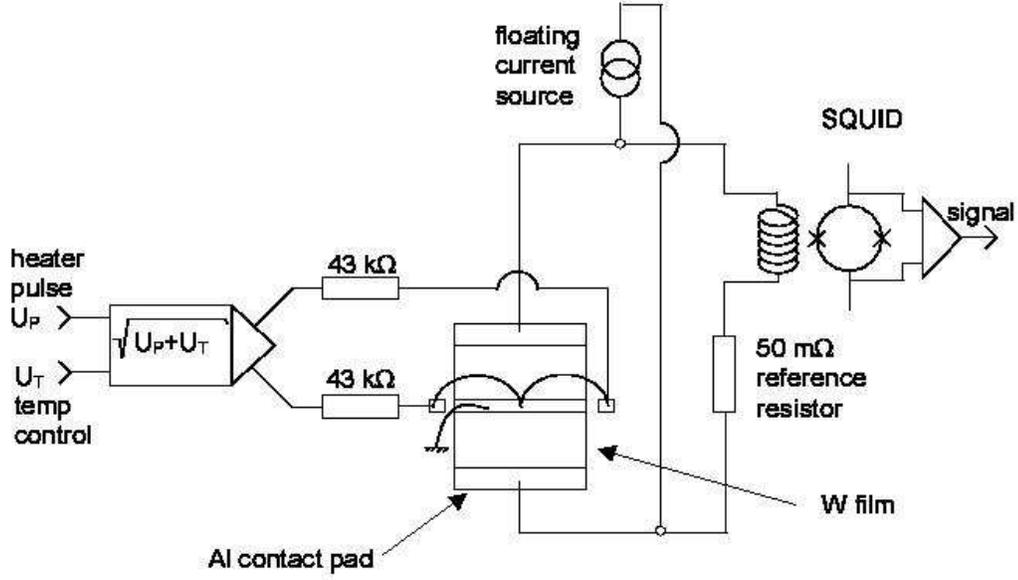,width=\textwidth,angle=0}
\end{center}
\vspace*{-11cm}
\caption[]{ Schematic thermal and electrical connections to the
thermometer in the phonon channel together with elements of 
the external readout and heater circuits. (See text) }
\label{fig-connect}
\end{figure}

The electrical and thermal connections of the thermometer are shown
in
Fig.~\ref{fig-connect}.  Thermal connection and grounding of the
detector is
provided by two gold bond wires of 25\,$\mu$m diameter, connecting
a
gold contact pad in the middle of the tungsten film with the copper
detector
holder. The holder in turn is thermally connected to the mixing
chamber of the
dilution refrigerator, which is stabilized at 6\,mK.  Electrical
connection to the read out circuit is made by superconducting
aluminum wires. 
These are bonded from the
aluminum pads on each end of the thermometer to electrically
insulated
contact pads on the copper holder. Superconducting wires connect
the
contact pads to the
SQUID readout circuit. These connections are screwed and not
soldered to avoid  the radioactivity of solder joints.

The resistance of the tungsten thermometer
($\sim$0.3\,$\Omega$) is
measured by means of a two-armed parallel circuit also shown in 
Fig.~\ref{fig-connect}. A bias current of a few $\mu A$
is supplied by a floating current source. One arm of the circuit is
given
by the superconducting film and  the other  consists of  the input
coil of a  SQUID in series with a reference resistor
(50~m$\Omega$),
which provides a sensitive measurement of current changes. The
SQUID is 
operated in flux locked loop mode. 
A rise in the thermometer resistance and so an increase in current
through
the SQUID coil is then observed as a rise  in
SQUID output voltage.

A heater for controlling the temperature of the thermometer and for
injecting the test pulses is provided by a 25\,$\mu$m diameter gold
wire bonded to a gold pad in the center of the 
thermometer and to small aluminum pads on either side of the
thermometer
(Fig.~\ref{fig-connect}). Superconducting aluminum bond wires, with
negligible thermal conductance at the low operating temperature,
 link these aluminum pads  to insulated pads on the copper holder
for 
electrical connection to the  heater circuit. The external heater
circuit
also shown in  Fig.~\ref{fig-connect}
sums the voltages U$_T$ for controlling the temperature and the 
voltage  U$_P$ from a pulser module and computes the square 
root of the sum so that the test pulse always deliver a known
heating proportional to U$_P$. The differential output of the
heater circuit  is connected via 43~k$\Omega$ series resistors to
the heater. 
 
\subsection{Light channel}

\begin{figure}[htb]
\begin{center}
\epsfig{file=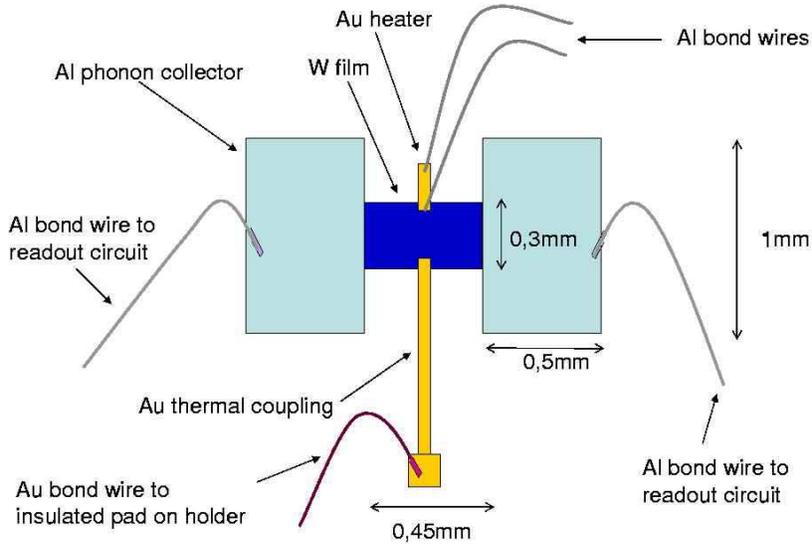, height=13cm,angle=-90}
\end{center}
\caption[]{Thermal and electrical connections to the sensor on the
light detector.  Aluminum/tungsten phonon
collectors surround the  tungsten  thermometer. Connections
shown are the  aluminum bond wires for electrical
connection to the SQUID and heater circuits and the  
gold bond wire for thermal
contact to the heat bath (copper holder).}
\label{fig-light-detector}
\end{figure}

At the operating temperature,  1\,\% or less of the energy
deposited in the CaWO$_4$ is detected as scintillation light. 
The cryogenic light detector used to detect these
scintillation photons consists of a silicon wafer ($30 \times 30
\times 0.4$~mm$^3$) with a tungsten thermometer. To improve the
efficiency for collecting  the energy of the few
photons, the thermometer is  supplemented by 
aluminum phonon collectors. This sensor, consisting of the 
thermometer and phonon collectors, is shown in 
Fig.~\ref{fig-light-detector}. In making these
sensors, an original tungsten film  is covered by a
superconducting 1$~\mu m$ thick aluminum layer on two sides
(0.5~mm$^2$ each), leaving  a small
uncovered tungsten film in the middle. This small uncovered region 
serves as the
thermometer.  The remaining aluminum/tungsten
bilayer is superconducting with a transition temperature close to
that
of pure aluminum (1.2 K) and so has a negligible heat capacity at
the operating temperature. 
These phonon collectors
absorb high frequency athermal phonons with an efficiency  similar
to that of the
tungsten thermometer, leading to the  creation of long lived
quasi-particles,
which in
turn efficiently deliver their energy to the tungsten
thermometer \cite{Loidl}. 
In this way we obtain a large collecting area while maintaining 
a thermometer of small heat capacity  $C$.

In CaWO$_4$ the time constant for the emission of the scintillation
photons is on the order of milliseconds at mK
temperatures~\cite{Philippe}. Thus   to integrate the energy of the
scintillation photons a  long thermal relaxation time
$\tau=C/G$ is
required for the light detector.  A sufficiently small thermal
coupling  $G$
of the
thermometer to the heat bath is obtained by a 50 nm thick, 
1.5$\times$0.1\,mm$^2$
 gold film overlapping the tungsten film at one end (lower end
in the figure \ref{fig-light-detector}), which is connected to the
heat bath
via a 25 $\mu$m gold wire bonded to an  electrically
insulated pad on the detector holder.  The electrical insulation of
this pad aids in
 the immunity of the detector against electronic
interference. This thermal coupling scheme gives the desired
thermal
relaxation
time on the order of
several milliseconds.  For the temperature regulation  and for the 
heating test
pulses, a second small gold film partially overlaps the
tungsten thermometer on the other end 
(upper end in figure \ref{fig-light-detector}). 
This film is connected via aluminum bond wires to two contact pads
on the
detector holder. Apart from the larger series resistors of
1~M$\Omega$ 
in the heater lines, heater and SQUID circuits for the light
channels are the same as for the phonon channels.

\subsection{Data acquisition}

For the data acquisition system
the output voltage of the SQUID electronics is split into two
branches. In 
one branch the pulse is shaped and AC-coupled to a trigger unit,
while in the other the signal is
passed
through an anti-aliasing filter and 
 DC-coupled to a 16-bit transient digitizer. 
The time base of the transient digitizer was chosen to 
be $40\,\mu$s, providing about 30 samples  for the rising
part  of the pulse in the phonon channel.
The total record length of 4096 time bins includes
a ``pre-trigger'' region of 1024 samples
to record the baseline before the event, while
 the ``post-trigger'' region of 3072 samples contains the pulse
itself. The
phonon and light channels were read out together, whenever one 
 or both  triggered. The transient digitizer data 
was then  written to disk for off-line analysis.
After each trigger there is a dead time of $\sim 70$\,ms to allow
for the readout and the sampling of the next pre-trigger region.
Pulses  from the other module which arrive within half
of the
post-trigger period 
are also recorded together with
 the time delay with respect to the first trigger. This
discarding 
of triggers in the other module arriving after half  of
the 
post trigger period contributes a dead time 
of 102.4 ms  per event for this other module. The accumulated
dead
time is measured with 
separate clocks for each channel. 

 In addition to  pulses from particle interactions, 
the data acquisition system  also
records the response to the  periodic test pulses which are
applied to each thermometer in order to monitor the
behavior of the detectors. 

\subsection{Temperature stability and energy calibration }
Although the width of the transition from superconducting to
normal resistance for a detector film is on the order of a mK,
for highly stable detector response the operating point on the
transition curve and so the temperature must be constant
 to within a few $\mu$K.  Short term temperature 
control was maintained by using the baseline of the SQUID output
voltage 
between pulses as the
temperature indicator and then regulating the current
to the heater on the thermometer. The baseline sampling
 rate was 10 Hz, and  during a pulse the sampling was
interrupted and the heater current kept constant.
These processes were run  under computer control
with a PI-algorithm capable of  recognizing pulses. This system
deals with flux quantum losses in the SQUID by not responding to
large ($>\phi_0/2$) jumps in SQUID baseline if they  occur very
quickly, namely in one
sampling period.

To take care of the very rare instances when a true temperature
jump
has taken place in a  short time, a 
  second independent control loop  provides for long term stability
by  checking for deviations from the desired operating point,  as
follows.
Every two seconds, large
heater pulses ($\sim$ 1.5 MeV equivalent in phonon channels),
which partially
saturate the detectors are injected. The energy of these pulses is 
large enough to drive the superconducting film 
near to the top of the transition, where it is almost
normal-conducting.
   The resulting SQUID output voltage will thus depend on how far
the
operating point was from the top of the transition and a possible
shift of the operating point can be recognized and corrected.
Hence for each thermometer there is  effectively a temperature
standard
given by the response of the film to a large heater pulse.
  The few  short periods with noticeable deviations
from the operating point are cut from the data.  In 
Table~\ref{tab:times} this is 
called the ``stability cut''.

\begin{figure}[tb]
\begin{center}
\epsfig{file=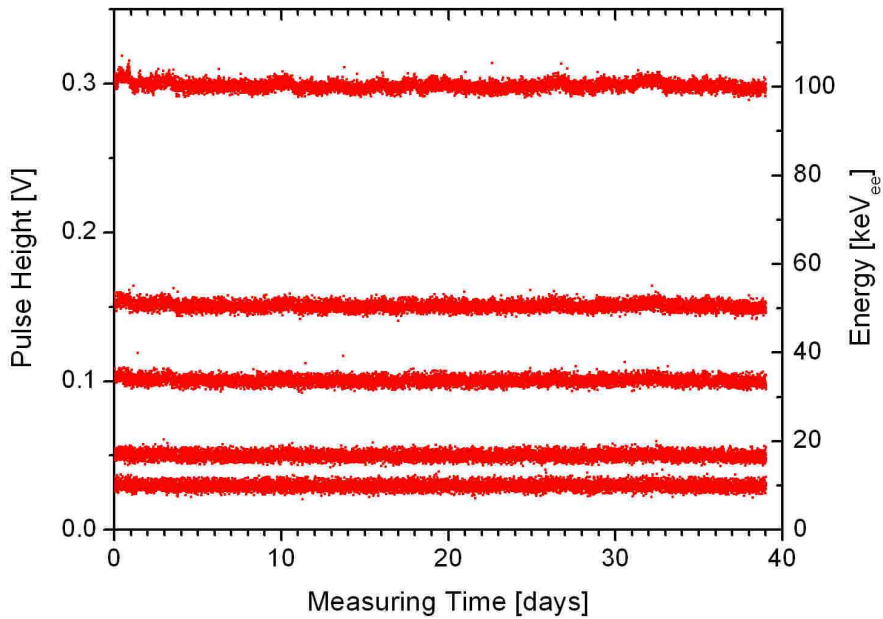, width=13cm}\end{center}
\caption[]{The measured pulse height for heater test pulses on
light detector BE13 as a function of time during the entire
dark  matter run. The energy scale on the right side refers to
effective    electron  energy in the CaWO$_4$ crystal (see
text). The actual  energy seen by the light detector is about 
a factor of 100 smaller. The detector is seen to be stable within
resolution.  }
\label{fig-ch4_stability}
\end{figure}

The heater used for  temperature regulation is also used
to inject the periodic heater test pulses
 monitoring  the long term stability of the energy calibration
and
testing the trigger efficiency close to  threshold.  These
heater
pulses are produced by exponential voltage pulses from a pulser
module, with the decay time adjusted to create detector pulses 
resembling true particle pulses. As previously remarked,  the
voltage for regulating the
temperature and the heater test pulses are added and passed through
an
analog square-root circuit to obtain a linear dependence of the
injected heater  energy vs. the amplitude of the pulser voltage.
The test pulses have a number of discrete energies spanning the
energy range of interest and were sent every 30\,s during both 
dark matter
and calibration runs. Fig.~\ref{fig-ch4_stability} shows the
stability
of the light detector BE 13 during the entire dark matter run.

The  energy corresponding to a pulse is determined by the pulse
height. This in turn
 is found by a template  fit procedure. 
For each detector a  template is made by averaging many measured
pulses of a given type, i.e. test pulses or pulses from 122~keV
$\gamma$'s 
from the calibration source in
the
case of the  CaWO$_4$ phonon detector. 
The pulse height of a given signal is then found by fitting  the
template  with an
amplitude scale factor, base line and onset time as free fit
parameters. The  amplitude scale factor then yields the pulse
height.
 Finally, a  calibration  translates the pulse height  into an
energy.     

The  CaWO$_4$ phonon detector, providing the energy of an
event, is calibrated with an external source
to give an absolute energy determination.  A $^{57}$Co source
(122~keV and
136~keV
$\gamma$ lines) is inserted into the shielding via a removable
plug,
illuminating the detector modules from below.  Comparison of the
pulse
heights of the 122~keV photons with heater pulses 
of similar energy provides an absolute
calibration for the voltage of the  heater pulses in terms 
of an equivalent $\gamma$ energy. 
The calibration
is transferred to the lower energies needed for the WIMP analysis
 by means of the heater test pulses.

\begin{figure}[tbh]
\begin{center}
\epsfig{file=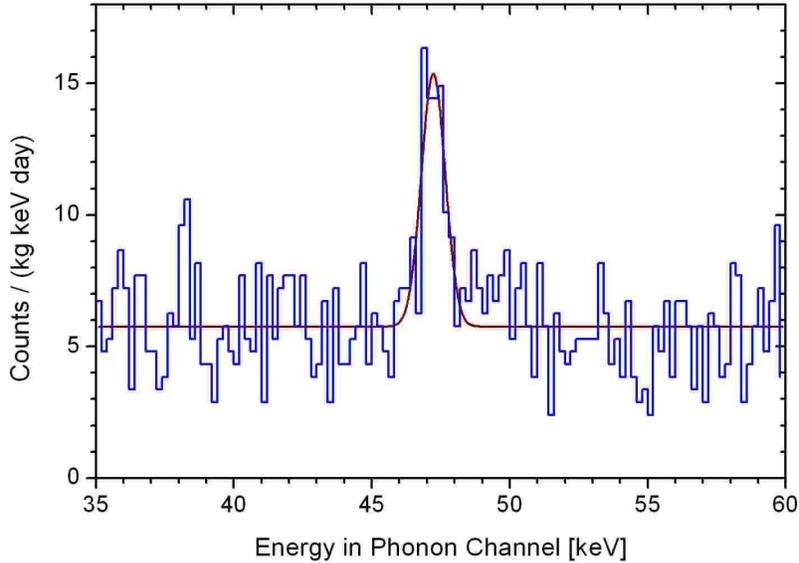, width=12cm,angle=0}
\end{center}
\caption[]{Peak with (3.2 $\pm$ 0.5) counts/day, of width 1.0\,keV
(FWHM) appearing in  the phonon channel (Daisy)
during the 53 day dark matter run. We attribute this peak to an 
$^{210}$Pb contamination  in the vicinity of the
detectors.}
\label{fig-daisy-pb210}
\end{figure}

The accuracy of the energy calibration  from 10 to 40~keV, as
relevant for the WIMP search, is in the range of a few percent.
This can be inferred from a peak at 47.1~keV shown in
Fig.~\ref{fig-daisy-pb210},
which appeared in
the energy spectrum of the phonon channel during
the dark
matter run with a rate of (3.2 $\pm$ 0.5) counts/day. 
If we associate this peak with the 46.54~keV
$\gamma$'s from an external
 $^{210}$Pb contamination, the calibration with the heater pulses
puts 
it 1.2\,\% too high. This accuracy of the calibration is
consistent 
with the accuracy 
 of about 1\,\% of the analog square-root circuit used to 
 inject the  heater pulses. 
The tendency
to slightly overestimate the energy of the events will put our dark
matter limits on the conservative side. The width of the 47~keV
peak is
1.0~keV (FWHM),   
 identical with that for the heater pulses. This good resolution
over 53 days again confirms 
the  stability of the response during the dark matter run.

For small
deposited energies
the temperature rise is considerably
smaller than the width of the film's superconducting transition and
so we obtain an
approximately
linear relation between pulse amplitude and energy up to at least
170~keV in all detector channels.  To deal with  larger pulses
 we have developed an new method where we  exclude that part of the
pulse  where
non-linearity sets in,  and then perform the fit on the thus
truncated 
pulse.  Essentially,  this means that large
pulse heights are determined by the duration of the
signal. We find this method quite effective and that the pulse
heights obtained from such a truncated fit are
approximately linear in energy up to several MeV.

\begin{figure}[tbh]
\begin{center}
\epsfig{file=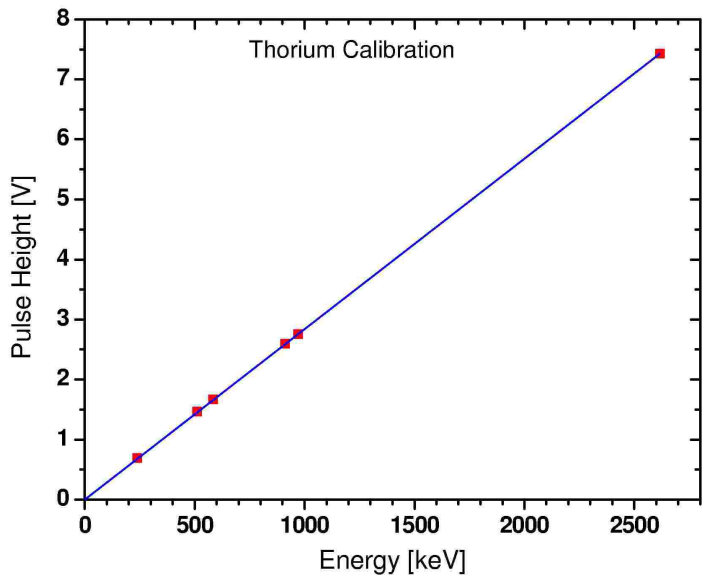, height=12cm,angle=-90}
\end{center}
\caption[]{Measured pulse height of the phonon channel (``Daisy'')
vs. energy for a set of $\gamma$ peaks from a $^{232}$Th
source. The pulse height was determined with a truncated
template
fit.  }
\label{fig-Th-calibration}
\end{figure}

Fig.~\ref{fig-Th-calibration} shows results obtained with a
$^{232}$Th
calibration source.  The pulse height obtained from the truncated
fit procedure
is plotted as a function of $\gamma$ energy for a number of
$\gamma$
peaks
of known energy. The straight line connecting the highest point
with
the origin serves to visualize the good linearity.  A refined
variant of
the truncated fit method was applied for a detailed study of the
$\alpha$
background of the detectors. These results are presented in a
separate paper \cite{Cristina}, where the $\alpha$ decay of
$^{180}$W  is definitively established for the first time. We note
that the low recoil energy
data used in the  present  dark matter analysis do  not involve
this truncation procedure.

The hardware threshold of the detectors was set to a recoil energy
of
about 5~keV, and 100\,\% trigger efficiency was confirmed
throughout
the
dark matter run by means of the lowest energy heater pulses,
corresponding to
8.5~keV and
10~keV in the  phonon channels for Julia and Daisy respectively. As
mentioned,
we obtain good
energy 
resolution: $\Delta$E = 1~keV (FWHM)
for 46.5~keV $\gamma$ rays and
$\Delta$E = 7~keV for 2.3~MeV $\alpha$'s. Typical pulse shape
parameters are
$\tau_{rise}$ = 1.1~ms for the rise time and $\tau_{dec}$ = 15~ms
for the fast
decay time.  At
energies
of a few tens of keV where the WIMP signal is expected, a
background
count
rate of $\sim$ 6 electron/photon events / (kg keV day), before
rejection
by the light signal, is obtained.

The  calibration of the light detectors follows the same procedure.
However, since we are not concerned with an absolute energy
determination, 
the calibration is provided by the
light output of the main detector with incident 122 keV $\gamma$'s.
Thus the pulse  height produced in the light detector with
122~keV photons incident on  the main detector is assigned the
nominal value (``electron equivalents -ee'') of 122~keV.
Given this normalization, the response of the light
detector thermometer may then be linearized by using  the heater
pulses applied to the thermometer, the heater pulses being 
adjusted to deposit  known fractions of the  energy necessary to
give the 122~keV pulse height.
There remains the question of a possible non-linear relation
between the energy deposit in the main detector and its light
output, as is
commonly observed in scintillators. This 
deviation from linearity would show up  as a curvature  of the
$\gamma$ band  in  Fig.~\ref{fig-daisy_2d}.
However, no curvature is evident.  More quantitatively, for the
Daisy/BE13 module,
 the light output with 10 keV and 40 keV in the main detector  is 
3.8~\%  and 2.8~\%   below the  expectation from
the above procedure of normalizing at 122~keV and linearizing with
the heater pulses.
 For the module Julia/BE14, which has a somewhat poorer light
resolution, these numbers are 9.4~\% and
6.7~\%.

\subsection{Energy resolution of the light channel}
\begin{figure}[b]
\begin{center}
\hspace*{-1cm}
\epsfig{file=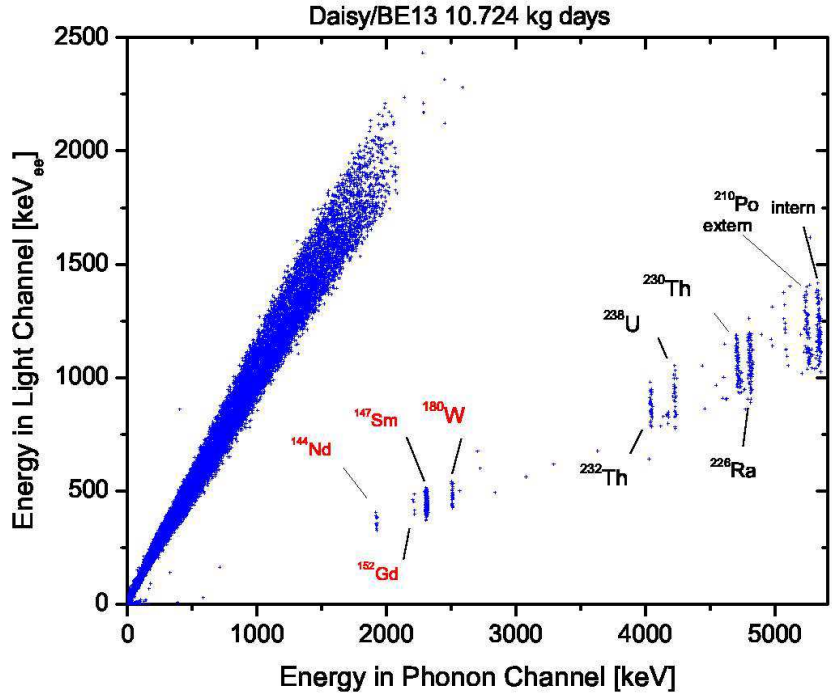, width=13cm}
\end{center}
\caption[]{ Energy in the light channel vs.\ energy in the phonon
channel recorded with the module Daisy/BE13. The uniform upper
band is from $\gamma$ and $\beta$ interactions in the CaWO$_4$
crystal, the lower band shows  a series of discrete $\alpha$ peaks, mostly
from the natural decay chains.}
\label{fig-daisy_2d}
\end{figure}

 The energy resolution of the light channel plays
an important role in the analysis. It reflects not only the
performance of the thermometer system but also fluctuations in
light production and collection.
In Fig.~\ref{fig-daisy_2d}  a scatter plot with energy in the light
channel vs. energy in the phonon channel over a wide energy
range  from 
Daisy/BE13 in the dark matter run is shown. Besides the 
continuous
$\gamma$ band a series of discrete $\alpha$ lines is visible in
a band with lower light yield. These  $\alpha$ peaks
are identified and discussed in detail in \cite{Cristina}.

An increasing width with energy of the $\gamma$ band, is seen  in
Fig.~\ref{fig-daisy_2d},
 arising from
an almost linear energy dependence of the light channel 
resolution. 
This may result from a slight
dependence
of the light output with respect to the position of the energy
deposition within
the 
CaWO$_4$ crystal. Therefore we use data
from the $^{232}$Th calibration to
extract the energy dependence of
the  resolution,   since in these data 
the low energy
continuum originates from Compton scattered high energy $\gamma$'s,
which interact uniformly in the volume of the crystal and so
are spatially distributed as 
expected
for 
neutron or WIMP-induced events. 
The resolution in the light channel for this thorium data was
modeled by a power series 
\begin{equation}
\Delta E = \sum_{i=0}^{3} A_{i} E^{i},    \label{eq:polynom} 
\end{equation}
for  energies between zero and 300\,keV,
 yielding the coefficients listed in Table
\ref{tab:light-resolution}. 
We observe that BE13 has the better
resolution at low energy as indicated by the smaller value of
$A_0$.

 The electronic noise of the better light detector
and hence its energy resolution when extrapolated to zero energy, 
is about 2\,keV (FWHM), slighly worse than  the energy resolution
of  1.5\,keV (FWHM) for  the 
heater pulses. This  compares 
favorably with photomultipliers, which  reach about the same 
performance with the best scintillating crystals of NaI(Tl). 
\begin{table}[tbh]
\caption{Polynomial coefficients describing the energy dependence
of the energy resolution of the light channels BE13 and BE14 in 
the energy range from 0 to 300\,keV.}

\begin{center}
\begin{tabular}{lcccc}
Detector & $A_0$ [keV$_{ee}$] & $A_1$  & $A_2$ [keV$_{ee}^{-1}$] 
& $A_3$[keV$_{ee}^{-2}$] \\\hline
BE13    & 1.974       & 0.24347   & -0.5794 10$^{-3}$    & 0.1368
10$^{-5}$ \\
BE14    & 3.446       & 0.24218   & -0.1617 10$^{-3}$    &-0.7127
10$^{-7}$ \\
\end{tabular}

\label{tab:light-resolution}
\end{center}
\end{table}

\subsection{Quenching factors} \label{sect:Qf}
 The determination of a nuclear recoil acceptance region 
in the phonon-light plane is based 
on a knowledge of the ``quenching factor'', i.e. the reduction
factor for  
the light output of a nuclear recoil event relative to an 
electron/photon event of the same energy.  Since this
plays an important role in our analysis, we present a short
discussion of our work on quenching factors.

 First, in the work of several years ago ~\cite{Meunier}, a
quenching factor of $Q$=7.4 was found for neutrons on  CaWO$_4$.
This was with an  apparatus similar to the one here with a 
cryogenic detector and neutrons from an $^{241}$Am-Be
source.  The recoil energies studied 
 covered the range  10 keV to 150 keV. 

  Since  a very  simple method of  pulse height determination was
used in ref.~\cite{Meunier}, we have reanalyzed this data
with the template fit procedure~\cite{spread}. Furthermore,
 the more recently developed truncated fit
method explained in section 2.4   allows us to  treat 
recoils  at higher energy.  We present the 
quenching factors found in this reanalysis for
 various energy ranges in
table~\ref{tab:neutron_quenching_factors}.
The  value of $Q$ found is marginally higher than the old
$Q$=7.4. 
\begin{table}[htb]
\begin{center}
\caption{Quenching factors measured with a cryogenic detector and
neutrons from an $^{241}$Am-Be source. From a reanalysis of the
data of ref.~\cite{Meunier}, with a  more sophisticated pulse
height determination. Statistical errors (1 $\sigma$ ) are given.} 
\vspace*{3mm}
\begin{tabular}{cc}
energy range [keV]& quenching factor (reanalysis)\\
\hline
10 to 20   & 10.0 $\pm$ 0.8    \\ 
20 to 40   &  8.9 $\pm$ 0.4  \\
40 to 100  &  8.2 $\pm$ 0.2   \\
100 to 150 &  8.2 $\pm$ 0.3   \\
\end{tabular}
\label{tab:neutron_quenching_factors}
\end{center}
\end{table}

 As a check on the reproducibility of the results of~\cite{Meunier}
a similar but independent test has been performed with  a  somewhat
larger CaWO$_4$ crystal, yielding essentially the same
results~\cite{stark}, with a small energy dependence and  the
same or slightly larger quenching factor.
In addition,
another group has reported an energy independent quenching factor
of $Q$=10 for CaWO$_4$ with  neutrons from a $^{252}$Cf source
\cite{Coron}. A further
subtlety is connected with the fact that an $^{241}$Am-Be source
has a harder energy spectrum than expected for Gran Sasso
neutrons \cite{Hesti}. This
would  indicate
the use of a somewhat smaller value of $Q$ for the Grand Sasso data,
where there would then be  less scattering on the heavier nuclei
than in the laboratory tests.  Now a larger  value of $Q$ in
analyzing the dark matter data leads to a
stronger rejection power.  Therefore,  in view of the various small
differences and uncertainties, in  the following we will take the
most conservative approach and use our original small value of
$Q$=7.4 to define nuclear recoils. We note however that variations of
$Q$ by
several units in the vicinity of 10  has little effect on our final
conclusions (see discusssion in sect \ref{accept}).

 The above points concerned $Q$ for neutron-induced recoils. However
it is possible and indeed expected that WIMPs and neutrons scatter
predominantly on different nuclei.  Monte Carlo
studies \cite{Hesti} show that neutrons with an energy spectrum
typical for Gran Sasso will produce recoils in the
energy range of
interest predominantly on the oxygen nuclei.  On the other
hand,
coherently
interacting WIMPs will scatter predominantly on the tungsten. It is
thus
necessary to know $Q$ for the  recoil of the different nuclei  in
CaWO$_4$, particularly tungsten. We have performed  dedicated
experiments for this purpose,  with 18 keV ions
of $W$ and other elements incident
on a  room
temperature CaWO$_4$  crystal~\cite{Jelena}. We find that the
quenching factors
show a
systematic increase throughout the periodic table, with 
the value $Q=40 \pm 4.5 $ for tungsten, which we shall use in our
analysis.
 These results will be
published shortly 
in a separate paper~\cite{Jelena}, and an extension to low
temperatures is now in progress.

\begin{figure}[htb]
\begin{center}
\hspace*{-1cm}
\epsfig{file=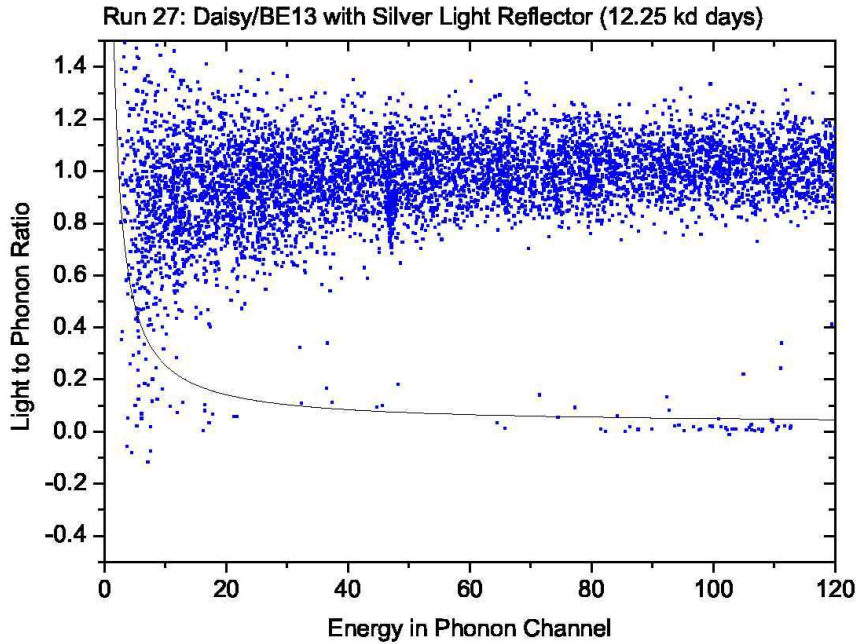, width=13cm,angle=0}
\end{center}
\caption[]{ Scatter plot showing a clustering of events from a
surface contamination   where a $^{210}$Po nucleus
decays into an $\alpha$ (5.4 MeV) which escapes and $^{206}$Pb (104
keV), seen  as a horizontal strip  between 80 and 110 keV  
slightly below the solid line. These
events may be used to find the quenching factor for Pb in the
experiment.  They were  later eliminated by using a reflector  foil
which also scintillates so that an alpha particle moving away from
the main crystal produces light, resulting in a ``high light"
event. The solid line is the 90\% boundary for $Q$=40 (see section
\ref{accept})}\label{1to65}
\end{figure}

However, we can also give direct low temperature evidence   in support
of $Q\approx 40$ for tungsten,  as follows. Before it was learned
how to reject them,  a class of events
was observed in earlier runs of the present apparatus, originating
from an external surface
contamination of $^{210}$Po. This nuclide  originates from Radon,
and is a common  contaminant on or slightly below surfaces, where
it can be implanted by  alpha decays.
 It decays
to $^{206}$Pb 
and an $\alpha$ particle.
We show data  from the earlier run27 in Fig.~\ref{1to65}.
The events in question are seen   as a well defined horizontal
strip between 80 and 110 keV, clustered
along and slightly below the solid line. They occur when
 a $^{210}$Po nucleus on the reflector foil surrounding the module
decays into the  $\alpha$ particle of 5.3 MeV and the $^{206}$Pb
nucleus  with 104 keV. If the $\alpha$  moves away from the
CaWO$_4$ detector and the $^{206}$Pb goes into the detector,  the
result is a low energy, low light, event. The energies of the
events extend below the 100~keV of the $^{206}$Pb nucleus since the
$^{210}$Po was implanted on the surface of the foil or somewhat
below it by previous alpha decays, and the $^{206}$Pb nucleus loses
some energy on the way out. A similar process can take place when
the 
$^{210}$Po is on or near the surface of the CaWO$_4$ crystal. In
this case it is the alpha that  loses some energy on the way out,
leading to energies somewhat above 100~keV.

 With these events  indentified as $^{206}$Pb, their light yield  
may be used
to find the quenching factor for Pb in the experiment. We note that
the quite horizontal character of the cluster indicates constancy
of the quenching factor with energy and   we find  
\begin{equation}\label{qpb}
 Q_{Pb}=48.7 \pm 7.1~~~~~~~~~~~~~~~~~~~ (7\,mK)\;. 
\end{equation}

  This value   at
low temperatures  and in the same apparatus, for Pb, which is not
far in
the periodic table from
$W$,  together with the systematic mass dependence seen in
the room temperature experiments, persuasively
 supports  our use of $Q=40$  for $W$ in the analysis to
follow.

These   $^{210}$Po  events were  finally  removed by  using a
reflector  foil
which also scintillates.  Alpha particle moving away from the main
crystal now produce light when hitting the foil  and so turn this
class of event  into ``high light" events which do not enter into
our region of interest. Hence this strip of events is absent from
the later runs we will present for the dark matter discussion.

We note in passing that we can  also use the $^{210}$Po process to
provide a further argument  that  the quenching
factor is the same for surface events and interior events in our
system.
In Fig.~\ref{fig-daisy_2d}  there are two nearby peaks for
$^{210}$Po decay, labeled  ``internal" and ``external". ``Internal"
events are  in the interior of the crystal and so show the full
energy of the  $^{210}$Po decay.   The somewhat lower energy of 
the ``external"  peak corresponds to surface events where the
$^{206}$Pb nucleus escapes and only the $\alpha$ remains in the
crystal. We find that the light yield is the same for both peaks, 
showing, a least for alpha particles, that the quenching factor is
the same for interior and surface events.
This supplements the argument in \cite{Meunier} that the absence of
any splitting of  the electron-photon band, where one has electrons
mostly near the surface and photons mostly in the interior,
demonstrates that bulk and surface events have the same quenching
factor. 
\section{Results and discussion}

In late 2002 and early 2003 a number of short runs were 
  undertaken to optimize  the prototype CRESST-II modules. We
report here on data
taken with two
detector
modules during the period from January 31 to March 23, 2004. Table
\ref{tab:times}
summarizes the measurement time, dead time, the time removed by the
stability cut  and the resulting total exposure.  
\begin{table}[htb]
\begin{center}
\caption{Measurement Times and Exposures}
\vspace*{3mm}
\begin{tabular}{lccccc}
module & measuring & dead time  & stability & mass  & total
exposure \\
        & time [days]& time [days] & cut [days]  &[grams]     & [kg
days]   
  
\\\hline
Julia/BE14    & 37.572       & 3.391   & 0.518  & 291.4   & 9.809 
     
\\
Daisy/BE13    & 39.043       & 3.469   & 0.621  & 306.8      &
10.724     
 \\
\end{tabular}
\label{tab:times}
\end{center}
\end{table}

Before this  dark matter run the detectors were calibrated with a 
$^{57}$Co source as described above and following
the  run  the $^{232}$Th test  reported in Fig.~\ref{fig-Th-calibration} 
was performed.

\subsection{Nuclear recoil acceptance region }\label{accept}
\begin{figure}[bp]
\vspace*{-1cm}
\begin{center}
\epsfig{file=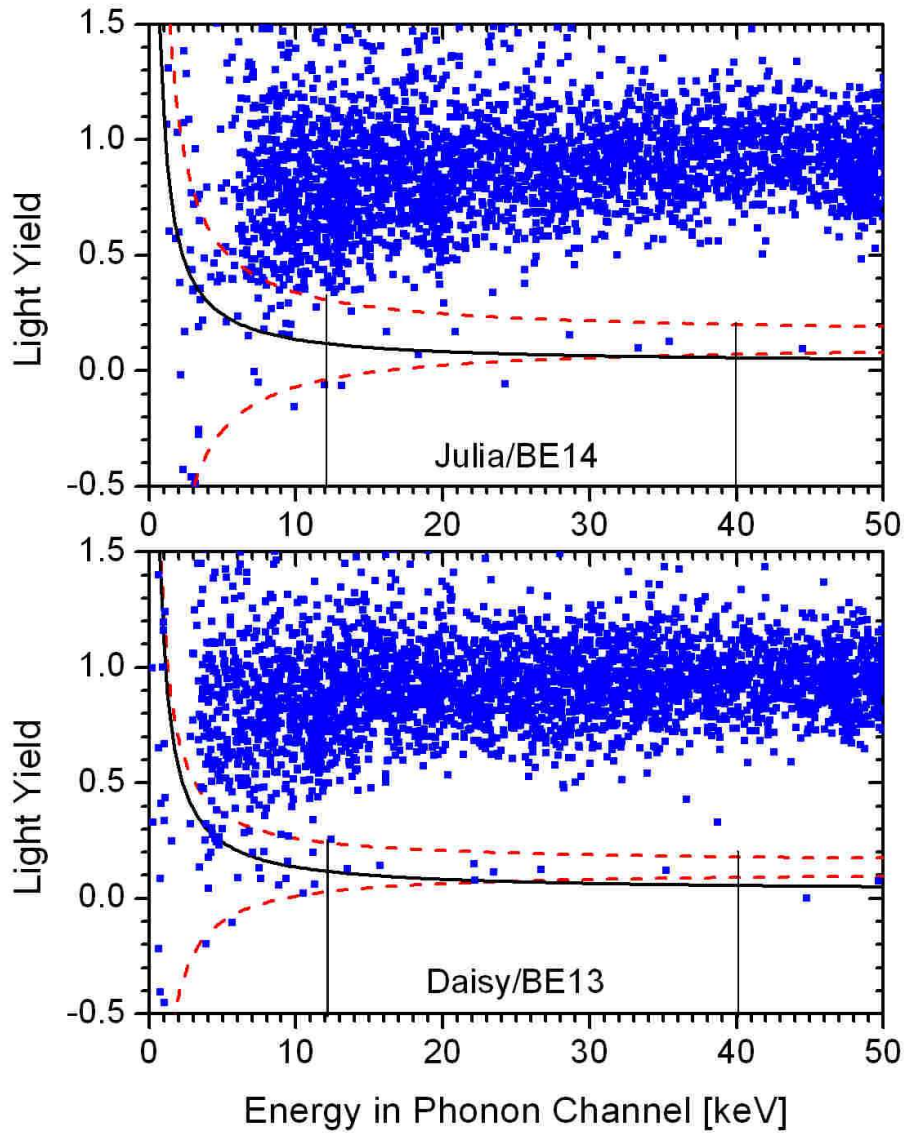, width=13cm,angle=0}
\end{center}
\caption[]{Low energy event distributions in the dark matter run 
for the two modules Julia/BE14 (top)  and Daisy/BE13 (bottom). The
vertical axis represents the light yield  expressed as the ratio
 (energy in the light channel/energy in the phonon channel), and  
 the horizontal axis the energy  in the phonon channel.
With a quenching factor of 7.4 for  nuclear recoils, the region
below the  upper  dashed curves will contain
90\,\%  of the nuclear recoils. The lower dashed curved shows the
10\,\% boundary with this quenching factor. 
Taking a quenching factor of 40 for tungsten,
the region below the solid curves will contain 90\,\% of  tungsten
recoils. The vertical lines indicate the energy range used 
for the WIMP analysis}
\label{fig-ratio_12_34}
\end{figure}

The data from the dark matter run is presented in
Fig.~\ref{fig-ratio_12_34}.
The  low energy region  for the two modules is shown in
   scatter plots where the vertical axis is the light to
phonon energy ratio, while the horizontal axis gives the phonon
 or total energy.
The points with negative energy  arise in the pulse fitting 
procedure  when a negative amplitude results  for small pulses
close to the noise.

The determination of a nuclear recoil acceptance region 
in the phonon-light plane is based on a knowledge of the quenching
factor.  
For this calculation we use the quenching factor $Q$=7.4,
as discussed in section~\ref{sect:Qf}.  
As mentioned there, somewhat higher values
have been obtained in other measurements, but  using the smaller
value of
$Q$=7.4 will put our dark matter limits on the conservative side.
 Using this quenching factor and  assuming 
Gaussian fluctuations parameterized with the 
energy resolution of the light and phonon channels, we obtain 
the dashed lines on the plots of Fig.~\ref{fig-ratio_12_34}.
 The upper dashed line gives the boundary below which 90\,\%  of
the
nuclear recoils are expected. We also show  
the  boundary below
which  10\,\% of the recoils for  $Q=7.4$ are expected. 
This is indicated by the 
 lower dashed lines.  Since  $Q=7.4$ refers to incident neutrons,
 the area between the  90\,\% and 10\,\%  lines will give an
impression of where the neutron background is expected to lie.
This 10\,\% line does not enter into the calculation of the WIMP
limits 
themselves, however.

 Since these calculations used the light channel resolutions from
Table 
\ref{tab:light-resolution},
determined with photons on the CaWO$_4$ detector, they also involve
the  assumption  that the  resolution of the
light channel does not depend on how the energy was deposited.  We
can
check this assumption by comparing the light channel resolution in
the two bands of Fig.~\ref{fig-daisy_2d} at the same value of  the
light output. If we do this at the nominal light output of
484\,keV$_{ee}$, 
corresponding to the $\alpha$ line at 2.31 MeV, we find
$64\pm 7$\,keV in the $\alpha$ band and $77 \pm 8$\,keV in the
electron/photon band; so within errors the resolutions are
consistent.

\begin{figure}[tb]
\begin{center}
\epsfig{file=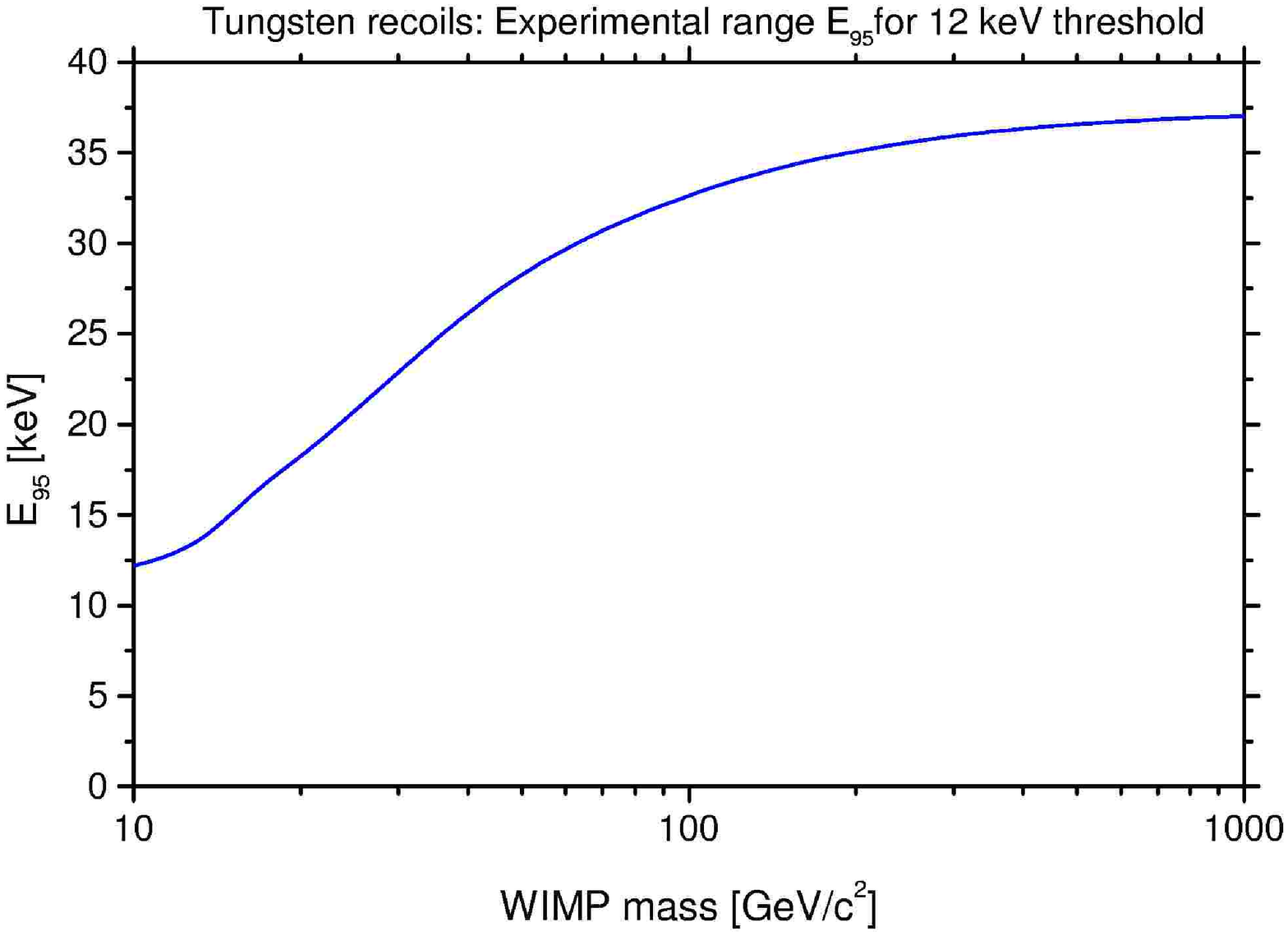, height=12cm,angle=-90}
\end{center}
\caption[]{ Recoil energy $E_{95}$ for WIMP-tungsten elastic
scattering as a function of
WIMP mass. $E_{95}$ is the recoil energy  below which 95\,\% of the
events above the threshold of 12~keV lie.  The WIMP flux
and nuclear form factor employed are
described in the text.}
\label{range}
\end{figure}

Finally,  an analysis
threshold of 12~keV was chosen in order to stay well above the
energy where 
electron and nuclear recoil bands intersect. The  upper energy for
the analysis was chosen to be  40~keV, since  due to the
suppression
of higher energy recoils by form factor effects, more than  95\,\%
of all WIMP
induced
tungsten recoils  are expected below this energy for any WIMP mass.

This is  shown in
Fig.~\ref{range}.
 Thus our nuclear recoil acceptance region on the plots of
Fig.~\ref{fig-ratio_12_34} are  below the upper  dashed lines   
and  from 12 to 40~keV in the phonon or total energy channel.

 In this acceptance region
there are a total of  7 (Daisy) and  9 (Julia) events for  the two
modules. 
Using the exposures of Table \ref{tab:times} and these 16 events we
obtain  
the rate
\begin{equation}\label{rate}
R(nuclear~ recoils)=(0.87\pm 0.22) /{\rm(kg\,day)} \; ,
\end{equation}
 for  12~keV $\leq$ $E_{recoil}\leq$ 40~keV.
 A 10\,\% acceptance correction has been applied to compensate for 
recoil events lost by the 90\,\% cut. The error in Eq.~\ref{rate}
represents simply
the statistical error. To check on systematic effects due to the
uncertainty of the quenching factor we find upon varying  $Q$ 
 that the number of recoils below the 90~\% acceptance
line 
for Daisy/BE13 remains the same 7 counts for quenching factors in 
the range from  
6.7 to 11.1.  For Julia/BE14 the result is stable for quenching
factors 
from 6 to 9.3.  Higher values of $Q$ will of course lead to stronger
limits. 
 
We note that most of these events lie  between the 90\,\% and
10\,\%
curves for $Q=7.4$ and so are in the region expected for
neutron-induced recoils. Furthermore,  
 Monte Carlo simulations for our setup
without  neutron shielding \cite{Hesti}
yield an estimate for the
 neutron background of about 0.6 events/(kg day) for 12~keV $\leq
E_{recoil}\leq$ 40~keV, in reasonable
agreement with Eq.~\ref{rate}. Both these remarks suggest  the
observed 
events are due to neutron background. 

 Double scattering events are possible for neutrons,
but   none of these  16  events  involved
coincidences 
between the two detector modules. However this may be understood as 
due to the fact that 
the detectors were mounted at a relatively large separation,   with
about 10~cm between  centers. We   estimate that if the  16
events were all  neutron events there should be less than 0.1
double scatters, consistent
with the  observation of none.  (Many  double Comptons are of
course seen in the electron/photon band.)

If we now nevertheless ignore the neutron background and  attribute
all 16 events to WIMP
interactions, we can set  a conservative upper limit for the WIMP 
scattering cross section.
 We calculate such a
 limit for coherent  (``spin independent'')
interactions,
 assuming a standard halo model (isothermal halo, WIMP
characteristic 
velocity: 220 km/s, mean Earth velocity: 232 km/s, 
local WIMP density: 0.3 GeV/c$^2$/cm$^3$).  
The Helm spin-independent form factor is used for the nucleus
\cite{Helm} with the
parameterization suggested in Ref.~\cite{Lewin} and A$^2$ scaling
for coherence.
The 
full line
in Fig.~\ref{fig-exclusion} shows this limit calculated with the
optimum
interval method of ref.~\cite{Yellin},
together with the DAMA (1-4) 3 $\sigma$ signal region ~\cite{DAMA}
and exclusion 
limits from EDELWEISS \cite{Edelweiss} and CDMS \cite{CDMS}.
Essentially identical
results are obtained when using the data of each module separately.

\subsection{Identification of the recoil nucleus}

\begin{figure}[b]
\begin{center}
\epsfig{file=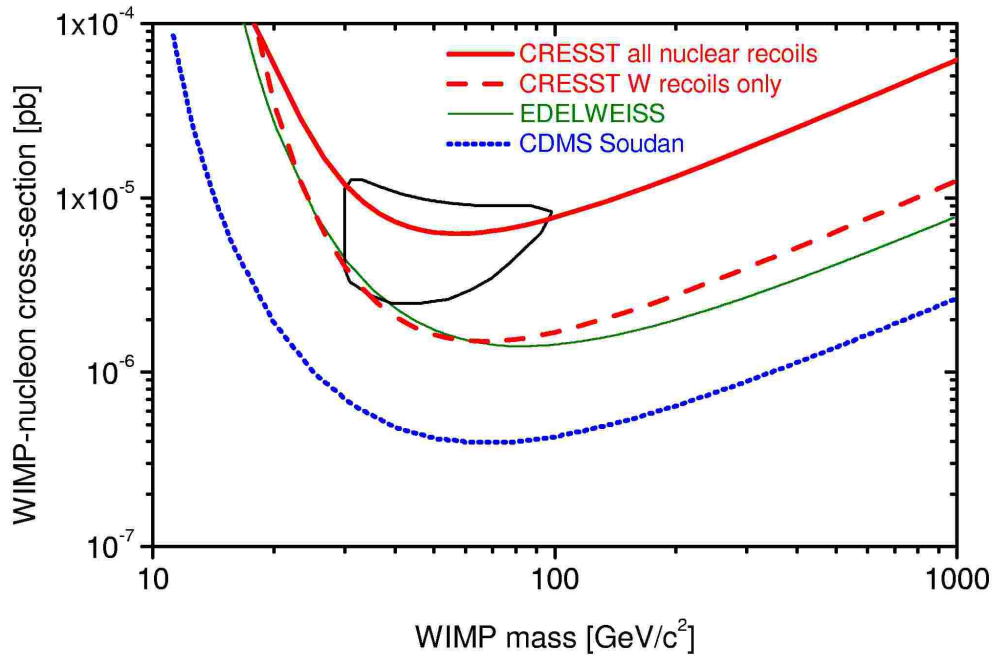, height=14cm,angle=-90}
\end{center}
\caption[]{
 Coherent  or spin independent scattering exclusion limits from the
dark matter run. 
 Using a  
 quenching factor of $Q=7.4$ to identify nuclear recoils as
explained in the text we obtain the full curve.
Using $Q=40$ to identify tungsten recoils  and the data of the
better module Daisy/BE13  give the thick dashed line. Regions
above the curves are
excluded
at 90\,\% CL. The enclosed region represents the claim of a
positive
signal by the DAMA collaboration~\cite{DAMA}. }
\label{fig-exclusion}
\end{figure}

In the above discussion we simply imposed a global limit on the
light yield to identify nuclear recoils.
Further important improvements in our technique are to be
anticipated
if it were possible to use the information from the phonon-light
output to identify which particular nucleus is recoiling. 
For neutrons in this recoil energy  range our Monte Carlo
simulation
shows that the dominant scattering  in 
the 
CaWO$_4$ is on the oxygen, with less than 5\,\% of the nuclear
recoils due to  tungsten. Hence the $Q$ value derived from the
neutron scattering
experiments is essentially
determined by oxygen recoils.  On the other hand for coherently
interacting WIMPs  the scattering is  dominated by the
tungsten, and a much higher value of $Q$ applies.

As discused in section~\ref{sect:Qf} we shall adopt $Q$ = 40 for
tungsten. With this value    we 
expect 90\,\% of the tungsten recoils to be below the
full lines in Fig.~\ref{fig-ratio_12_34}. 
One of the two modules (bottom  in
Fig.~\ref{fig-ratio_12_34}) has no  recoil events  below 
this full line in the energy range from 12 to 40~keV, while the
other module has 3 events.

This difference between the two modules corresponds to the better
resolution in the light channel for the module with no
events. The better resolution  will inhibit ``leakage'' of
neutron-induced recoils into  the region of the tungsten
events. This may occur, for example,   when
 a statistical fluctuation
in the light channel produces a low light value for a
neutron-induced recoil. We can check the plausibility of this
explanation with an estimate of the ``leakage'': using the 
 resolutions determined for the light channel and
 Gaussian fluctuations we calculate the probability, for each
event in the neutron band, that it would have appeared below the
tungsten line. This procedure leads to the  following estimate  
 for the ``leakage":
for the module Daisy/BE13, 1.1 events; and for the module
Julia/Be14,
3.6 events. Most of this leakage occurs at the lower energies
where the tungsten acceptance line (solid line) lies above the
10\,\%
boundary for the neutron recoils. 
If  we were to move the analysis threshold from 12 to
25~keV  the leakage would become much smaller, 0.092
and
0.62 events, respectively. ``Leakage'' thus does appear to give a 
plausible explanation for the difference in the behavior of the two
modules.

  We now use  the better module  Daisy/BE13, where there are no
tungsten
events, to set a WIMP limit. We 
obtain   the thick dashed line in
Fig.~\ref{fig-exclusion}. To check systematic effects due to choice
of the  acceptance region  we can lower the threshold  
to 10~keV to include the two events   
at 10.5~keV and 11.3~keV below the tungsten line for Daisy/BE13 
in Fig.~\ref{fig-ratio_12_34}. This leads to
 essentially the same curve on the exclusion plot and
confirms the 
stability of the results against a variation of the analysis
threshold.

As a second check on systematics we can lower the quenching factor
for tungsten--that is weaken the rejection power of the light
channel. The number of events 
below the solid curve in Fig.~\ref{fig-ratio_12_34} (Daisy) remains
at zero for quenching factors down to $Q$ = 35.8, where an event at
22.2 keV enters. 
This single event would move the minimum of the exclusion curve 
from 1.6$\times$10$^{-6}$~pb to 2.3$\times$10$^{-6}$~pb .
The result is then stable until $Q$=23.4, where a second event at
13.5 keV enters.
 This would move the minimum of the exclusion curve to 
3.8$\times$10$^{-6}$ pb.  This confirms the stability of the
results at about $10^{-6}$~pb  against 
uncertainties in the value of $Q$ for tungsten. 

At a WIMP mass of 60~GeV/c$^2$ these tungsten limits, 
obtained without neutron 
shielding, are identical to the limits set by EDELWEISS
\cite{Edelweiss} also assuming coherent interactions. 
Recent results from CDMS at the Soudan Underground Laboratory
\cite{CDMS} have
improved these limits by a further factor of four.

\section{Summary and Perspectives}

The present results were obtained without a neutron shield and
the WIMP sensitivity appears to be limited by the neutron
background when all nuclear recoil events in the CaWO$_4$ are
naively
attributed to WIMP interactions. However, we expect that the
background neutrons, scattering predominantly on the light
elements of the detector,  lead to events distinguishable by light
yield
from coherently scattering WIMPs, which scatter predominantly on
the tungsten.  Thus  if we exploit the  high
quenching factor for tungsten,  the tungsten recoils can be
separated. On this basis
we  have no tungsten recoils for the better of the two modules in
the acceptance range of 12 to
40~keV.  For the efficiency of this separation, the
resolution of 
the light channel is crucial and the difference in the behavior
of the two modules can be traced to different resolutions
in the light channel.

We observe that the coherent scattering hypothesis enters doubly in
our analysis. Once, as for all types of detectors, in enhancing
the rate on a heavy nucleus,  given a certain WIMP-nucleon cross
section. But secondly in a manner particular to the present method
in that 
it leads to  distinctly different anticipated light yields for 
neutron
background and WIMP-induced recoils.

Finally, we note that the method of identifying the recoil
nucleus by the location of the signal in the phonon-light plane
seems very promising, not only for the suppression of background
but also for the verification of a possible positive WIMP
signal~\cite{proposal} and for unraveling the quantum numbers of a 
WIMP candidate~\cite{pok}. To this end further study of quenching
factors and improvement of the light detectors are being actively
pursued.

\section{Acknowledgements}

This work was partially supported by 
the DFG SFB 375 ``Teilchen-Astrophysik'',
the EU Network ``Cryogenic Detectors'' (contract ERBFMRXCT980167),
the EU Network HPRN-CT-2002-00322  ``Applied Cryodetectors'', 
BMBF, PPARC, and two EU Marie Curie Fellowships.


\begin{thebibliography}{99}
\bibitem{coh}
 In general, coherent scattering  will be characterized
by the square of a linear combination of proton and neutron
numbers. However we will follow the common practice and use $\sim
A^2$.
\bibitem{spread}
 There might appear to be a question as to the constancy of $Q$  at 
low recoil energy  since in Fig.~3b of  ref.~\cite{Meunier} the
light
yield 
appears to  increase at the lowest energies.
However, this is an artifact of
the very  simple method of pulse height determination used in
ref.~\cite{Meunier}.
The pulse heights were just taken from
the 
highest point of the pulse. With this algorithm an upward
 noise fluctuation near the peak of  the pulse will essentially
always  make the apparent light yield higher, while a downward
noise fluctuation will have little or no effect. With 
the template fit
procedure this asymetery is removed and we find  the
apparent upturn disappears.
\end{thebibliography}
\end{document}